# Price as a matter of choice and families of finitely–additive probabilities

Yaroslav Ivanenko

Banque de France[1]

**Abstract**    A version of indifference valuation of a European call option is proposed that includes statistical regularities of nonstochastic randomness. Classical relations (forward contract value and Black-Scholes formula) are obtained as particular cases. We show that in the general case of nonstochastic randomness the minimal expected profit of uncovered European option position is always negative. A version of delta hedge is proposed.

**Key words**   Derivatives Valuation, Decision Theory, Finitely-additive probabilities

## 1. Introduction

In spite of its almost perfect features, mathematical finance, with its focus on arbitrage and theory of stochastic processes (Harrison and Kreps (1979), Delbaen and Shechermayer (2010)), faces today several issues. One such issue is the omnipresent nature of statistical instabilities in financial markets and inability of contemporary stochastic models to take it into account (see, for example, Mandelbrot (2006), Epstein et al (2000)). Another issue is best exemplified in Back et al (1991), where the authors showed that countably–additive measures are insufficient in order to represent no arbitrage situations when the state space is infinite and that finite-additivity is required instead. Perhaps, these two issues are related. Perhaps, counatble-additivity, or stochasticity, seems to be there only to make a full use of contemporary theory of stochastic processes. That probability theory has some limitations, and that therefore it is important to distinguish between stochastic randomness and randomness in a broad sense was stressed already by Kolmogorov (1986). A positive answer on the above questions was given in Ivanenko (1990). It was shown that a closed family of finitely-additive probability distributions – called statistical regularity - describes a random in a broad sense phenomenon, Nonstochastically random stands for a synonym of statistically unstable. In particular, this formalism covers realizations of statistically unstable random phenomena for any state space.

[1] **Disclaimer: the ideas expressed and the results obtained in this article reflect the views of the author and are not necessarily shared by the institution to which the author is affiliated**

The present article explores one of the possible ways of introduction of this remarkable result in mathematical finance.

Beginning from von Neumann and Savage, decision theory has provided a framework to many topics in mathematical economics (Mas-Collel et al (1995)) and mathematical finance (see, for instance, Markowitz (1959)). Some formulations of decision theory, with its focus on uncertainty and representation of preferences of economic agents, require existence of set functions - families of finitely-additive probabilities and capacities (see, for instance, Ivanenko (1986), Gilboa (1989), Shmeidler (1989)) – that, prior to the result mentioned above, were difficult to interpret from a statistical standpoint. This is one of the reasons why these theories have not found yet their way into mathematical finance. The only version of decision theory that has established a place in finance is expected utility theory. Contemporary indifference pricing methodology (Carmona (2008)) can be seen as an application of expected utility theory to the problems of asset valuation. The most interesting results of this exploration concern the link between indifference pricing and arbitrage, or risk-neutral pricing (see for instance, Bellini et al (2002)). Perhaps, the absence of arbitrage opportunities does not require countable-additivity of probability distributions. Perhaps market incompleteness and nonstochastic randomness are two complementary features of the financial system. These questions seem to be still open and are important for further research in the field. It can be that considering abovementioned versions of decision theory in the context of indifference pricing can help answering these questions.

The present article considers in the context of indifference pricing the formulation of decision theory operating with closed families of finitely-additive distributions (Ivanenko 2010). Namely, we show how families of finitely-additive probability distributions enter valuation expression of a European option. This expression shows how the abovementioned nonstochastic randomness could be effectively priced by the market. We directly model the price of financial contract as a decision, as an object of choice. This modeling technique has strong practical justification. Moreover, unlike standard indifference pricing framework, the approach proposed here does not require the use of concave utility functions in order to model the phenomenon of risk aversion. The concavity feature responsible for risk or uncertainty aversion is in the shape of the decision criterion and one can use linear utility or loss functions, which is more appealing from practical viewpoint as well. One can see parallels of this approach and the arbitrage pricing theory proposed in Ross (1976, 1977). It seems as well that the approach proposed here is a natural framework for static replication arguments (see Derman and Taleb (2008). The described framework seems to have a lot in common with de Finetti ideas (1937), with price becoming a synonym of provision. Besides, de Finetti works are characterized by weakening of the countable-additivity of probabilities.

In what follows we present our framework for one underlying and for an arbitrary portfolio (Section 2.1-2.2). Then we show (Section 2.3) how with the help of the put-call-parity-like argument of indifference some classical results of mathematical finance (forward contract value and Black-Scholes formula) are obtained as particular cases. In Section 2.4 an expression of delta hedge is proposed for statistical regularities of general form.

## 2. Price as decision and nonstochastic randomness

### 2.1. One underlying

Let $f(\theta)$ be a pay-off of a financial contract, depending on the value of uncontrolled (unknown) parameter $\theta$. For example, $f(\theta) = (\theta - \theta^*)^+$, where $\theta$ is the price of the underlying financial variable at the maturity date $T$, $\theta^*$ is the strike. Let at the moment $t=0$ we observe the transaction of the pay-off $f(\theta)$ at a certain price $u$. Let the buyer finances the purchase with a loan, so that by the time $T$ she would owe the bank the sum $ue^{rT}$. Let the seller places the proceeds of the sale in the bank, so that by the time $T$ the bank will owe him the sum $ue^{rT}$. We shall write that by the moment $T$ the profit and loss of the buyer and the seller will be, respectively,

$$L_b(\theta, u) = -ue^{rT} + f(\theta) \qquad (1)$$

and

$$L_s(\theta, u) = +ue^{rT} - f(\theta). \qquad (2)$$

Our goal is to interpret the observed price $u$ of the financial contract as a decision of the buyer and of the seller. Let $U = \{u : u \in \mathbb{R}\}$ be the set of decisions, or actions, that in this case are prices. Let $\Theta$ be the set of values of the random parameter $\theta$. Note that the values of the above profit and loss functions $L$ represent outcomes with the natural preference relation on them ("the bigger the profit, the better") that coincides with the order of real numbers.

Following the framework of *decision systems,* we construct two sets, $Z_b = (\Theta, U, L_b(.,.))$ and $Z_s = (\Theta, U, L_s(.,.))$, called *decision scheme* of the buyer and, respectively, of the seller.

We admit further that the buyer and the seller belong to the *class $\Pi_1$ of decision makers*, that is those whose preference relation on actions is represented by means of the criterion $L_Z^*: U \to \mathbb{R}$ that has the form

$$L_Z^*(u) = \max_{p \in P} \int L(\theta, u) p(d\theta), \qquad (3)$$

where $Z = (\Theta, U, L(.,.))$ is the *decision scheme, $U$* - is the set of decisions $u$, $\Theta$ - is the set of values of the uncontrolled (random) parameter $\theta$, $L: \Theta \times U \to \mathbb{R}$ - is a loss function, and where $P$ is a so called *statistical regularity* in the form of a closed set of *finitely-additive* probability measures on $\Theta$. Note that the sets $\Theta$ and $U$ are *arbitrary* and the function $L$ is *bounded*. These circumstances will allow us to extend the set of decisions in a way that suits most the goal of the article. Any statistical regularity $P$ describes a statistically unstable *random* phenomenon, called *nonstochastically random*[2]. If $L$ has the meaning of a utility function, as in (2.1)-(2.2), the criterion (2.3) of maximal expected losses becomes the criterion of minimal expected utility[3]

$$L_Z^*(u) = \min_{p \in P} \int L(\theta, u) p(d\theta). \qquad (3')$$

In what follows we shall use the form (3') of the criterion and call it the criterion of minimal expected profits.

We admit further that the buyer and the seller not only belong to the same class $\Pi_1$ of decision makers, but share as well the same view on the statistical regularity $P$ of the behavior of the uncontrolled random parameter $\theta \epsilon \Theta$. In our case this means that the price $u$ of the transaction is that decision $u$ that makes the buyer and the seller *indifferent* between the profit and loss profiles $L_b(\theta, u)$ of the buyer and $L_s(\theta, u)$ of the seller. In other words, we can interpret the fact of the transaction at the price $u$ as equality[4]

$$L^*_{Z_b}(u) = L^*_{Z_s}(u),\qquad(4)$$

or, substituting (3')

$$\min_{p \in \boldsymbol{P}} \int L_b(\theta, u) p(d\theta) = \min_{p \in \boldsymbol{P}} \int L_s(\theta, u) p(d\theta).\qquad(5)$$

---

[2] Unlike stochastically random phenomena, *nonstochastically random* phenomena are those random phenomena that do not possess the property of stability of frequencies. A model of realization of such a phenomenon is constructed without usual probabilistic assumptions and is called *sampling directedness,* or *directedness in the space of samples*. The theorem that states that any sampling directedness has a statistical regularity, and any statistical regularity corresponds to a sampling directedness is proved in Ivanenko (1990, 2010). Note that the words *directedness, net* and *generalized sequence* are synonyms.

[3] This is achieved by means of the change of sign of the loss function. In this connection see as well De Groot (1970), Ivanenko (1986), Gilboa (1989) and the Appendix.

Taking into account (1)-(2), and due to boundedness of $\int f(\theta) p(d\theta)$, we obtain

$$- u \, e^{rT} + \min_{p \in \boldsymbol{P}} \int f(\theta) p(d\theta) = u \, e^{rT} - \max_{p \in \boldsymbol{P}} \int f(\theta) p(d\theta).\qquad(6)$$

Thus the observed price $u$ may be interpreted as the average of the following type

$$u = e^{-rT} \; \frac{\min\limits_{p \in \boldsymbol{P}} \int f(\theta) p(d\theta) + \max\limits_{p \in \boldsymbol{P}} \int f(\theta) p(d\theta)}{2},\qquad(7)$$

where $\boldsymbol{P}$ is a statistical regularity on $\Theta$. The formula (7) means that market negotiates, or agrees upon, a particular choice of the statistical regularity $\boldsymbol{P}$. A stochastic analogue of this is known as *implied distribution*.

Remark though that if one requires that $L^*_{Z_b}(u) = 0$ and $L^*_{Z_s}(u) = 0$ then one obtains, respectively, that

$$u_b = e^{-rT} \; \min_{p \in \boldsymbol{P}} \int f(\theta) p(d\theta)\qquad(7')$$

that can be interpreted as the *bid price* and

$$u_s = e^{-rT} \; \max_{p \in \boldsymbol{P}} \int f(\theta) p(d\theta)\qquad(7'')$$

that can be interpreted as the *ask price.* The indifference price (7) can be interpreted thus as the *mid price.*

In the case of complete uncertainty, when $\boldsymbol{P}$ is a the set of all finitely-additive probability measures on $\Theta$, including all Dirac delta distributions, the formula (7) becomes

---

[4] When the buyer and the seller belong to different classes of decision makers and do not share the views on randomness, the equality of this type would not make sense. What is, probably, more important, the buyer and the seller may have very different asset-liability and financing constraints, resulting in diversity of the profit and loss functions *L*.

$$u = e^{-rT} \; \frac{\min\limits_{\theta \in \Theta} f(\theta) + \max\limits_{\theta \in \Theta} f(\theta)}{2}. \tag{8}$$

In the case of stochastic character of the statistical regularity, that is when the set $P$ contains a single stochastic, that is countably-additive probability measure $p$, the formula (7) becomes

$$u = e^{-rT} \; \int f(\theta) p(d\theta) \tag{9}.$$

Because of the equality (4) the price (7) may be considered as another version of the *indifference price*: see Carmona (2008). In our case, however, a decision maker is indifferent between the role of the buyer and the role of the seller. Here indifference means that she has no grounds to prefer one operation to another.

In statistical literature, for example in De Groot (1970), the criterion of the type (3) is called *risk*. Therefore our indifference price may be interpreted as that decision that makes equal the risks of the counterparties. This interpretation leads to a parallel with the valuation theory developed in Ross (1976, 1978).

### 2.2. Arbitrary portfolio

Since we admit that decision makers belong to one class of decision makers, it is reasonable to rewrite the above for the case of a single decision maker as well as for the case of multiple securities. Let us first consider only one underlying. Let the random parameter $\theta$ be the price of the underlying security by the future time moment $T$, and let $\Theta$ be the set of its possible values. Let $F = \{ \Theta \to \mathbb{R} \}$ be a set of all real bounded functions on $\Theta$ that we interpret as pay-offs of financial instruments depending on the random parameter $\theta \in \Theta$. Let $q \in Q = \{q : q \in \mathbb{R}, |q| < \infty\}$ represent the sense and the amount of the operation ($q > 0$ for purchase, $q < 0$ for sale, $q = 0$ for no operation. Let $U = \{u : u \in \mathbb{R}, |u| < \infty\}$ be the set of prices. Then let $D = F \times Q \times U$ be the set of decisions. This means that a decision maker chooses three elements: a contract, her role and the price. Supposing that the decision maker finances her purchases via a bank account and places the proceeds of sales there as well, the profit and loss function $L : \Theta \times D \to \mathbb{R}$ will be

$$L(\theta, d) = q\left(-u e^{rT} + f(\theta)\right), \;\; d = \left(q, u, f(.)\right) \in D, \;\; f \in F. \tag{10}$$

In this way one constructs the *matrix decision scheme* $Z = (\Theta, D, L(.,.))$, an essential element of decision system. Since the representation (3) is valid for *arbitrary* sets $D$ and $\Theta$, and for *bounded* function $L$ one can extend the above setting (which is a model of a one period investment in a singular market) to the case of an arbitrary portfolio. Indeed, let there exist a collection of $j=1, \ldots, M$ underlying securities. Namely let $\theta = \{\theta_1, \ldots, \theta_M\}$ be their prices by the time moment $T$, and let $\Theta = \{\Theta_1, \ldots, \Theta_M\}$ be the collection of sets of their possible values. Without loss of generality we can consider $\Theta$ as the set of the states of Nature. Let $D^{(\infty)} = \cup_{n=1}^{\infty} D^n, D^n = \underbrace{D \times D \ldots \times D}_{n}$, where $D$ is the elementary decision set from (10). Then the profit and loss function $L: \Theta \times D^{(\infty)} \to \mathbb{R}$ has the form

$$L(\theta, d) = \sum_{i=1}^{N_d} q_i\big(-u_i e^{rT} + f_i(\theta)\big), \qquad (10')$$

where $d = \big(d_1, \ldots, d_{N_d}\big) \in D^{(\infty)}, d_i = (f_i(.), u_i, q_i) \in D, i = 1, \ldots, N_d, \theta \in \Theta$. In this case the matrix decision scheme takes the form

$$Z = \Big(\Theta, D^{(\infty)}, L(., .)\Big). \qquad (11)$$

Any decision scheme

$$Z' = \Big(\Theta' \subseteq \Theta, D' \subseteq D^{(\infty)}, L(., .)\Big). \qquad (11'')$$

is called a *market*. When for each $d_i \in d, i = 1, \ldots, N_d, u_i$ are fixed, we say that we deal with an *investment decision problem*; when $f_i \in d$ are fixed we say that we deal with a *valuation decision problem*; when neither is fixed we may say that we deal with a *market decision problem*. The set of values of the profit and loss function (10') comprises the set of outcomes with the natural preference relation on it: the bigger the profit, the better. Due to the nature of financial markets, *all* the elements introduced above are considered as bounded: $M < \infty, \forall d \ N_d < \infty, \forall j = 1, \ldots, M \ \inf \Theta_j < \infty, \sup \Theta_j < \infty$.

As in the previous section, her preference relation on actions is represented by means of the criterion (3') that now is written as

$$L_Z^*(d) = \min_{p \in \boldsymbol{P}} \int L(\theta, d) p(d\theta), \qquad (12)$$

where now $\theta \in \Theta, d = \big(d_1, \ldots, d_{N_d}\big) \in D^{(\infty)}, L(\theta, d)$ as in (10) and where $\boldsymbol{P}$ is a statistical regularity on $\Theta$. If for two decisions $d_1, d_2 \in D^{(\infty)}$ the decision maker has

$$L_Z^*(d_1) = L_Z^*(d_2), \tag{13}$$

then she has no grounds to prefer one decision to another, she, in other words, is indifferent with respect to the choice between them.

Remark though, that when $d_1 \in D^1, q = 1$, that is for the buyer, one can, requiring $L_Z^*(d_1) = 0$, conclude that

$$u = e^{-rT} \min_{p \in \boldsymbol{P}} \int f(\theta) p(d\theta) \tag{14'}$$

and interpret it as the *bid* price. Respectively when $q = -1$, that is for the seller,

$$u = e^{-rT} \max_{p \in \boldsymbol{P}} \int f(\theta) p(d\theta). \tag{14''}$$

and interpret it as the *ask* price.

In what follows we consider valuation decision problems for *M=1*. Now the question is what statistical regularity $\boldsymbol{P}$ will the decision makers use? We show in the next section how the argument of indifference leads to the analogue of a risk neutral measure, but in the case of nonstochastic randomness. In what follows relations (3')-(5) and (10)-(13) are our basic tool.

### 2.3. European options, forward contract and put-call parity: conditions on $\boldsymbol{P}$.

The pay-off of a European call option is $f(\theta) = (\theta - \theta^*)^+$, where $\theta$ – is the underlying stock price at the maturity date *T*, $\theta^*$ - is the strike,. Hence, using (2.7) we have

$$u_c = e^{-rT} \frac{\min_{p \in \boldsymbol{P}} \int (\theta - \theta^*)^+ p(d\theta) + \max_{p \in \boldsymbol{P}} \int (\theta - \theta^*)^+ p(d\theta)}{2}. \tag{15}$$

The pay-off of the corresponding European put is $f(\theta) = (\theta^* - \theta)^+$. Hence

$$u_p = e^{-rT} \frac{\min_{p \in \boldsymbol{P}} \int (\theta^* - \theta)^+ p(d\theta) + \max_{p \in \boldsymbol{P}} \int (\theta^* - \theta)^+ p(d\theta)}{2}. \tag{16}$$

For the corresponding forward contract on the underlying stock $f(\theta) = \theta - \theta^*$ . Hence

$$u_f = e^{-rT} \frac{\min_{p \in \boldsymbol{P}} \int (\theta - \theta^*) p(d\theta) + \max_{p \in \boldsymbol{P}} \int (\theta - \theta^*) p(d\theta)}{2} =$$

$$= e^{-rT} \left[ \frac{\min\limits_{p \in P} \int \theta p(d\theta) + \max\limits_{p \in P} \int \theta p(d\theta)}{2} - \theta^* \right]. \tag{17}$$

What meaning could we attribute to the first term in the brackets? Let us try to define as decision the forward price $\theta_F$ for the stock supposing that its current spot price is $\theta_0$. On one hand, $\theta_F$ must be such as to guarantee indifference between the long and the short forward positions, i.e. when, respectively,

$$L_l(\theta, \theta_F) = -\theta_0 e^{rT} + \theta_F,$$

$$L_s(\theta, \theta_F) = +\theta_0 e^{rT} - \theta_F$$

then one must have

$$\min\limits_{p \in P} \int L_l(\theta, \theta_F) p(d\theta) = \min\limits_{p \in P} \int L_s(\theta, \theta_F) p(d\theta). \tag{18}$$

Since $L_l(\theta, \theta_F)$ and $L_s(\theta, \theta_F)$ are point pay-offs (they do not depend on $\theta$), then

$$\theta_F = \theta_0 e^{rT}, \tag{19}$$

which is the classical forward price for a stock whose current price is $\theta_0$ (see Hull (2005)). On the other hand, in the case of non-deliverable forward, the indifference between long and short positions, the pay-offs of which in this case are, respectively,

$$L_l(\theta, \theta_F) = -\theta_F + \theta$$

$$L_s(\theta, \theta_F) = +\theta_F - \theta,$$

implies

$$\min\limits_{p \in P} \int L_l(\theta, \theta_F) p(d\theta) = \min\limits_{p \in P} \int L_s(\theta, \theta_F) p(d\theta), \tag{20}$$

where again $\theta$ is the stock price at the date $T$ and $\theta_F$ is the decision with respect to the forward price. Whence

$$\theta_F = \frac{\min\limits_{p \in P} \int \theta p(d\theta) + \max\limits_{p \in P} \int \theta p(d\theta)}{2}. \tag{21}$$

That is the average of the type (21), where $\theta$ - is the stock price at the future time moment $T$, $\Theta$ - is the set of its possible values, and $\textbf{\textit{P}}$ - is a statistical regularity on $\Theta$, has the meaning of the forward price as well. At the same time, from (19), we must have

$$\theta_0 = e^{-rT} \; \frac{\min\limits_{p \in \textbf{\textit{P}}} \int \theta p(d\theta) + \max\limits_{p \in \textbf{\textit{P}}} \int \theta p(d\theta)}{2}. \tag{22}$$

Provided the price $\theta_0$ is known, equation (22) becomes a condition on the "fair" statistical regularity $\textbf{\textit{P}}$ on $\Theta$.

Note that we would obtain (22) considering as well the following decision problem. Let today's stock price $\theta_0$ be a decision in the situation where a decision maker chooses between long and short stock positions held to a certain time horizon and financed with a bank account. Namely, let the profit and loss function of the long and short positions be respectively

$$L_l(\theta, \theta_0) = \theta - \theta_0 e^{rT}$$

and

$$L_s(\theta, \theta_0) = \theta_0 e^{rT} - \theta.$$

Being indifferent in this case means

$$\min_{p \in \textbf{\textit{P}}} \int (\theta - \theta_0 e^{rT}) p(d\theta) = \min_{p \in \textbf{\textit{P}}} \int (\theta_0 e^{rT} - \theta) p(d\theta) \tag{23}$$

or

$$\theta_0 = \; e^{-rT} \; \frac{\min\limits_{p \in \textbf{\textit{P}}} \int \theta p(d\theta) + \max\limits_{p \in \textbf{\textit{P}}} \int \theta p(d\theta)}{2}. \tag{24}$$

One can thus conclude that the choice of the spot price $\theta_0$ becomes a condition on $\textbf{\textit{P}}$ similar to (22).

Substituting (22), or (24), in (17) one obtains

$$u_f = \; e^{-rT} \; (\; \theta_0 e^{rT} \; - \theta^*) = \theta_0 - \theta^* e^{-rT}, \tag{25}$$

where $\theta_0$ – is the current stock price and $\theta^*$ - the strike. This quantity is usually called *the value of the forward contract*: see, for instance, Hull (2005). In other words, the problem of choice of the forward contact value is equivalent to the problem of choice of the spot price.

Now it is easy to retrieve the classical put-call parity relation. Consider the following decision scheme. On one hand, a decision maker can buy the call at the price $u_c$ (15), and simultaneously sell the put for the price $u_p$ (16). On the other hand, the decision maker can take a forward position, the value of which is (25). The profit and loss from holding long call-short put position is

$$L_{cp}\big(\theta, -u_c + u_p\big) = (-u_c + u_p)e^{rT} + (\theta - \theta^*)^+ - (\theta^* - \theta)^+ =$$

$$= (-u_c + u_p)e^{rT} + (\theta - \theta^*). \tag{26}$$

And the profit and loss from holding a long forward position is

$$L_f\big(\theta, u_f\big) = -u_f e^{rT} + (\theta - \theta^*). \tag{27}$$

According to (12) and (13), being indifferent between these payoffs means

$$\min_{p\in\boldsymbol{P}} \int L_{cp}\big(\theta, -u_c + u_p\big)p(d\theta) = \min_{p\in\boldsymbol{P}} \int L_f\big(\theta, u_f\big)p(d\theta), \tag{28}$$

or, taking into account (26) and (27),

$$u_c - u_p = u_f. \tag{29}$$

Substituting (15, 16, 25) in (29) we have:

$$\frac{\min_{p\in\boldsymbol{P}} \int (\theta - \theta^*)^+ p(d\theta) + \max_{p\in\boldsymbol{P}} \int (\theta - \theta^*)^+ p(d\theta)}{2}$$

$$- \frac{\min_{p\in\boldsymbol{P}} \int (\theta^* - \theta)^+ p(d\theta) + \max_{p\in\boldsymbol{P}} \int (\theta^* - \theta)^+ p(d\theta)}{2} =$$

$$= \theta_0 e^{rT} - \theta^*. \tag{30}$$

Provided the choice of the current stock price $\theta_0$ has been made by the market participants, this condition on statistical regularity $\boldsymbol{P}(\boldsymbol{\Theta})$ is the nonstochastic analogue of the no-arbitrage condition on the pricing measure of the usual stochastic case. Indeed, in stochastic case we would have from (30), as well as from (22) and (24),

$$\int \theta p(d\theta) = \theta_0 e^{rT}, \tag{31}$$

a condition that locks in the expectation of the stock price and yields the Black-Scholes pricing formula for a log-normal $p(\theta)$ with a volatility parameter σ.

### 2.4. Generalized delta

In the general case of statistical regularity $\boldsymbol{P}$ the minimal expected profits (3') of the uncovered European option position with pay-off $f(\theta)$ are equal for the buyer and the seller, and are always negative. Substituting (7) back into (5) one has

$$L_{Z_s}^*(u^*) = L_{Z_b}^*(u^*) = \min_{p \in \boldsymbol{P}} \int L_b(\theta, u^*) p(d\theta_T) = \min_{p \in \boldsymbol{P}} \int f(\theta) p(d\theta) - u^* e^{rT} =$$

$$= \min_{p \in \boldsymbol{P}} \int f(\theta) p(d\theta) - \frac{\min_{p \in \boldsymbol{P}} \int f(\theta) p(d\theta) + \max_{p \in \boldsymbol{P}} \int f(\theta) p(d\theta)}{2} =$$

$$= \frac{\min_{p \in \boldsymbol{P}} \int f(\theta) p(d\theta) - \max_{p \in \boldsymbol{P}} \int f(\theta) p(d\theta)}{2} \leq 0. \tag{32}$$

When statistical regularity $\boldsymbol{P}$ is stochastic, the minimal expected pay-off of the uncovered option position is zero. Relation (32) shows that the use of stochastic probability measures may be the reason why the markets systematically underestimate the risks of financial transactions.

It is important to note that it is not yet clear if in the case of statistical regularities of the general form the *dynamic hedging* framework is theoretically possible. First of all, we simply do not know yet how to describe the time evolution of statistical regularity. Second, from the valuation view point, the delta hedge construction is not necessary in order to determine the price of the call option: the price is given by (15) and (24). Third, from risk management perspective the knowledge of the delta is important, but even having obtained the pricing relations, the derivation of the option price sensitivity to the change of the underlying price is not trivial.

Nevertheless we can define a *static* hedge, or, as one may call it, a *generalized* delta. Indeed let us compose a portfolio of a European option and of a position in the underlying and require that its minimal expected pay-off (12) were not negative. Represent this portfolio construction as a problem of choice (11). Let $M = 1, \Theta = \Theta_1, d = (d_1, d_2) \in D \times D = D^2 \in D^{(\infty)}, d_1 = (f(.), 1, u^*), d_2 = (g(.), \delta, \theta_0)$, where $f(\theta)$ is as above, $g(\theta) = \theta$ in order to represent a position on the underlying itself, $u^*$ is chosen as (7), $\theta_0$ is chosen as (22) and $\delta \in Q$. Then, according to (10'),

$$L(\theta, d) = \sum_{i=1}^2 q_i\left(-u_i e^{rT} + f_i(\theta)\right) = -u^* e^{rT} + f(\theta) + \delta(-\theta_0 e^{rT} + \theta), \qquad (33)$$

and the decision scheme is given as $Z = (\Theta, D^2, L(,))$. The criterion (12) now is

$$L_Z^*(d) = \min_{p \in P} \int L \ (\theta, d) p(d\theta) =$$

$$= \min_{p \in P}\left(\left(\int f(\theta) p(d\theta) - u^* e^{rT}\right) + \delta\left(-\theta_0 e^{rT} + \int \theta p(d\theta)\right)\right) \geq$$

$$\geq \min_{p \in P}\left(\int f(\theta) p(d\theta) - u^* e^{rT}\right) + \delta \min_{p \in P}\left(-\theta_0 e^{rT} + \int \theta p(d\theta)\right) =$$

$$= \min_{p \in P} \int f(\theta) p(d\theta) - u^* e^{rT} + \delta\left(-\theta_0 e^{rT} + \min_{p \in P} \int \theta p(d\theta)\right) =$$

$$= \frac{\min\limits_{p \in P} \int f(\theta) p(d\theta) - \max\limits_{p \in P} \int f(\theta) p(d\theta)}{2} + \delta \frac{\min\limits_{p \in P} \int \theta p(d\theta) - \max\limits_{p \in P} \int \theta p(d\theta)}{2}.$$

Requiring the last sum be equal zero, one obtains

$$\delta = -\frac{\max\limits_{p \in P} \int f(\theta) p(d\theta) - \min\limits_{p \in P} \int f(\theta) p(d\theta)}{\max\limits_{p \in P} \int \theta p(d\theta) - \min\limits_{p \in P} \int \theta p(d\theta)}. \qquad (34)$$

This quantity of the underlying makes the minimal expected pay-off of the portfolio $d = (d_1, d_2) \in D \times D \in D^{(\infty)}$ non-negative. Are there conditions that make (34) converge to the standard Black-Scholes delta, $\delta \to \delta_{BS}$ ? What does (34) converge to when $P = \mu, \forall p \in P$, that is when random phenomena is $\mu$ - stochastic in the sense of the Definition 4.8 from the book Ivanenko (2010)? When $P = \mu$, does $\delta = \frac{\mu(\partial f)}{\mu(\partial \theta)}$? These questions are important because statistical regularities of the general form correspond to statistically unstable random phenomena, and hence, one is tempted to say, more real, than stochastic processes. This is yet another argument in order to question the dynamic hedging framework as it is done in Derman and Taleb (2008).

### 3. Discussion

The theoretical elements presented in this article are an adaptation of the decision theory described in Ivanenko (2010) to the problem of valuation of derivative contracts, a traditional problem of mathematical finance. This adaptation is an interpretation of observed phenomenon, namely of an observed transaction price. However, financial decision makers may belong to different classes of decision makers, not necessarily to the class $\Pi_1$, and thus may use very different criteria for their actions, if any. They may have as well very different views on the type of behavior of random variables. And transactions may still take place. The benefit from being a member of this class is obvious: one is prepared, in a manner of speaking, from the onset to statistically unstable random outcomes. The representatives of the class $\Pi_0$, that is those who follow the guidance of the expected utility criterion, are devoid of this benefit.

It seems that considering pricing problem as a problem of choice, or a decision, is a natural framework for static replication argument, as presented in Derman and Taleb (2008), and allows for a coherent introduction of the concept of nonstochastic randomness in mathematical finance.

In conclusion we would like to stress that due to Ivanenko (1990, 2010) the families of finitely-additive probabilities, that appear in the context of decision theory (see Ivanenko (1986), Gilboa (1989)) but that have never been accepted in mathematical finance, have finally acquired their statistical meaning: they describe statistically unstable, or nonstochastic, random phenomena. Therefore it seems reasonable to suggest that further exploration of statistical regularities of nonstochastic randomness, besides being a new research topic, may happen to be a road away from current underestimating of risks of financial transactions as well as one more argument in favor of those who disbelieve stochastic character of financial variables.

### Appendix
### On the definition of the class $\Pi_1$

We argue that decision makers belonging to the class $\Pi_1$ are those who would systematically prefer neutral or fully diversified investment strategy to a directional one. Namely, Condition 3 of the following definition, called in Ivanenko (1986, 2010) the *guaranteed result principle* generalized for mass events and reflecting *uncertainty aversion* of the decision maker, can be interpreted as the diversification argument, encouraging an investor to pursue neutral strategies. Below we reproduce the axiomatic definition of the class $\Pi_1$ and its characterization theorem from Ivanenko (1986, 2010).

**<u>Definition A1</u>** Let $\mathbb{Z}$ be the class of all ordered triples of the form $Z = (\Theta, U, L)$, where $\Theta, U$ are arbitrary nonempty sets and $L: \Theta \times U \to \mathbb{R}$ is a real bounded function. The triple $Z$ is called a decision scheme. We denote by $\mathbb{Z}(\Theta)$ the subclass of all decision schemes of the form $Z = (\Theta, ., .)$, where the set $\Theta$ is fixed.

**<u>Definition A2</u>** We define a *criterion choice rule* to be any mapping $\pi$, defined on $\mathbb{Z}(\Theta)$ and associating to every scheme $Z = (\Theta, U, L)$ some real function $L_Z^*(\cdot)$, *a criterion*, determined on $U$. We denote the class of all criterion choice rules by $\Pi(\Theta)$ and include in the subclass $\Pi_1(\Theta) \subset \Pi(\Theta)$ all criterion choice rules that satisfy the following three conditions:

C1. If $Z_i = (\Theta, U_i, L_i) \in \mathbb{Z}(\Theta), i = 1,2, U_1 \subset U_2$, and $L_1(\theta, u) = L_2(\theta, u) \ \forall u \in U_1, \forall \theta \in \Theta$, then $L_{Z_1}^*(u) = L_{Z_2}^*(u) \ \forall u \in U_1$.

C2. If $Z = (\theta, U, L) \in \mathbb{Z}(\Theta)$, $u_1, u_2 \in U$, then if $L(\theta, u_1) \leq L(\theta, u_2), \forall \theta \in \Theta$, then $L_Z^*(u_1) \leq L_Z^*(u_2)$, and if $a, b \in \mathbb{R}, a \geq 0$ and $L(\theta, u_1) = aL(\theta, u_2) + b, \ \forall \theta \in \Theta$, then $L_Z^*(u_1) = aL_Z^*(u_2) + b$.

C3. If $Z = (\theta, U, L) \in \mathbb{Z}(\Theta)$, $u_1, u_2, u_3 \in U$ and $L(\theta, u_1) + L(\theta, u_2) = 2L(\theta, u_3) \ \forall \theta \in \Theta$, then $2L_Z^*(u_3) \leq L_Z^*(u_1) + L_Z^*(u_2)$.

The next Theorem (which is a simplified version of Theorem 1 from Ivanenko (1986) or Theorem 5.2 from Ivanenko (2010) establishes the link between the properties of $L_Z^*(\cdot)$ and its structure.

**<u>Theorem</u>** Criterion $L_Z^*(\cdot)$ possesses the properties C1-C3 if and only if it has the following structure

$$(A1) \qquad\qquad L_Z^*(u) = \max_{p \in P} \int L(\theta, u) p(d\theta),$$

where $P$ is a statistical regularity on $\Theta$ in the form of a closed family of finitely-additive probability measures.

Note that in the above definition and theorem the sets $U, \Theta$ are arbitrary nonempty sets and the loss function $L: U \times \Theta \to \mathbb{R}$ is bounded.

If instead of the loss function $L$ one considers profit and loss function $\hat{L} = -L$, as we do in this article, then condition C3 of the above Definition becomes

C3'. If $Z = (\theta, U, \hat{L}) \in \mathbb{Z}(\Theta)$, $u_1, u_2, u_3 \in U$ and $\hat{L}(\theta, u_1) + \hat{L}(\theta, u_2) = 2\hat{L}(\theta, u_3) \ \forall \theta \in \Theta$, then $2L_Z^*(u_3) \geq L_Z^*(u_1) + L_Z^*(u_2)$.

The criterion *(A1)* of maximal expected losses then becomes the criterion of minimal expected utility (see Ivanenko (1986), Gilboa (1989))

$$(A1^{'}) \qquad\qquad L_Z^*(u) = \min_{p \in \boldsymbol{P}} \ \int \hat{L}(\theta, u) p(d\theta).$$

Below we demonstrate that if a decision maker uses criterion (A1'), then Condition 3 can be interpreted as the diversification argument (see as well Ivanenko (2010), page 111).

Let $F = \{\Theta \to \mathbb{R}\}$ be the set of all bounded real functions on $\Theta$ that we interpret as pay-offs of financial instruments, depending on a random parameter $\theta \in \Theta$[5]. Let $U = \{u : u \in \mathbb{R}\}$ be the set of prices of the pay-offs $F$. Let $q \in Q = \{-1, 0, +1\}$ represent the sense of the operation, -1 for a sale, +1 for a purchase and 0 for the absence of operation. Then let $D = F \times Q \times U$ be the set of decisions. This means that a decision maker chooses three elements: a contract, her role and the price. Supposing that the decision maker finances her purchases via a bank account and places the proceeds of sales there as well, the profit and loss function will be

$$(A2) \qquad\qquad L(\theta, d) = q\big(-u e^{rT} + f(\theta)\big), d = \big(q, u, f(.)\big) \in D, f \in F.$$

In this way one constructs in the matrix decision scheme $Z = (\Theta, D, L(.,.))$, an essential element of decision system.

Let

$$(A3) \qquad\qquad d_1 = (+1; f_1(.); u_1), \quad L(\theta, d_1) = -u_1 e^{rT} + f_1(\theta),$$

$$(A4) \qquad\qquad d_2 = (+1; f_2(.); u_2), \quad L(\theta, d_2) = -u_2 e^{rT} + f_2(\theta),$$

$$(A5) \qquad\qquad d_3 = \left(+1; \frac{f_1(.) + f_2(.)}{2}; \frac{u_1 + u_2}{2}\right),$$

$$L(\theta, d_3) = -\frac{u_1 + u_2}{2} e^{rT} + \frac{f_1(\theta) + f_2(\theta)}{2}.$$

Then it is obvious that

$$(A6) \qquad\qquad L(\theta, d_1) + L(\theta, d_2) = 2L(\theta, d_3), \forall \theta \in \Theta.$$

Provided

---

[5] Generally speaking, the elements of $F$ are the elements of the Banach space of all real bounded functions on $\Theta$.

$(A7)$ $\qquad L_Z^*(d) = \min_{p \in P} \int L(\theta, d) p(d\theta),$

show that

$(A8)$ $\qquad L_Z^*(d_1) + L_Z^*(d_2) \leq 2 L_Z^*(d_3).$

Indeed,

$(A9)$ $\qquad L_Z^*(d_1) = \min_{p \in P} \int L(\theta, d_1) p(d\theta) = -u_1 e^{rT} + \min_{p \in P} \int f_1(\theta) p(d\theta),$

$(A10)$ $\qquad L_Z^*(d_2) = \min_{p \in P} \int L(\theta, d_2) p(d\theta) = -u_2 e^{rT} + \min_{p \in P} \int f_2(\theta) p(d\theta),$

$(A11)$ $\qquad L_Z^*(d_3) = \min_{p \in P} \int L(\theta, d_3) p(d\theta) =$

$\qquad\qquad = -\dfrac{u_1 + u_2}{2} e^{rT} + \min_{p \in P} \int \dfrac{f_1(\theta) + f_2(\theta)}{2} p(d\theta).$

Since

$(A12)$ $\qquad \min_{p \in P} \int f_1(\theta) p(d\theta) + \min_{p \in P} \int f_2(\theta) p(d\theta) \leq \min_{p \in P} \int (f_1(\theta) + f_2(\theta)) p(d\theta).$

we obtain (A8).

This confirms that to use criterion *(A1)*, or *(A1')*, in order to estimate decisions ex ante and to prefer in situations of uncertainty neutral or fully diversified strategies are equivalent.